\documentclass[fleqn,twoside]{article}
\usepackage{espcrc2,psfig}

\newlength{\figwidth}
\newcommand{\simgt}{\stackrel{>}{\scriptstyle\sim}}

\title{Anisotropic lattices for precision computations
in heavy flavor physics%
       \thanks{Poster presented by H. Matsufuru}}

\author{
Hideo Matsufuru\address{%
    Yukawa Institute for Theoretical Physics, Kyoto University,
    Kyoto 606-8502, Japan \vspace{-0.25cm} },
Masanori Okawa\address{%
    Department of Physics, Hiroshima University,
    Higashi-hiroshima 739-8526, Japan \vspace{-0.25cm}},
Tetsuya~Onogi$^{\rm a}$
and
Takashi~Umeda$^{\rm a}$ }

\begin{document}

\begin{abstract}
We study the anisotropic lattice QCD
for precision computations of heavy-light matrix elements.
Our previous study in which the lattices are calibrated with a few
percent accuracy has already given results comparable to
the existing calculations.
This suggests that even higher precision may be achieved
by a more precise calibration of anisotropic lattices.
We describe our strategy to tune the gauge and quark parameters
with accuracies much less than 1 \% in the quenched approximation.
\end{abstract}

\maketitle

\section{Introduction}
  \label{sec:introduction}

Recent experimental progress at $B$ factories 
suggests that precise computation of hadronic 
matrix elements in lattice QCD is a key 
to the search for signals of new physics in flavor physics.
However, HQET or the relativistic approaches
still suffer from perturbative and/or discretization errors, 
which are typically of order of 10\%.
We therefore need yet another framework of the heavy 
quark in which one should be able to
(i) take the continuum limit,
(ii) compute the parameters in the action and the 
     operators nonperturbatively,
(iii) and compute the matrix elements with a modest computational
      cost.

As a candidate of framework which fulfills these conditions,
we investigate the anisotropic lattice on which the temporal lattice
spacing $a_\tau$ is finer than the spatial one $a_\sigma$
\cite{Aniso01a,Aniso01b,Aniso02a}.
The anisotropic lattice approach evidently satisfies above conditions
(i) and (iii).
Our expectation is that on anisotropic lattices the mass dependence
of the parameters becomes so mild that one can adopt
coefficients determined nonperturbatively at massless limit.
For precise computations of heavy-light matrix elements,
we also need to control all the systematic errors in the
extrapolations to the continuum limit.
Whether these promises will be practically satisfied should be
examined numerically.

So far we have investigated the feasibility of the approach
in the quenched approximation and with the tree-level tadpole
improvement for the $O(a)$ improved Wilson quark action
\cite{Aniso01a,Aniso01b,Aniso02a,Aniso02_fD}.
As will be summarized in Sec.~\ref{sec:stage1},
the results have been encouraging for further development
in this direction.
Therefore we have started the second stage of the study of
the anisotropic lattice for heavy quarks.
In this stage, we perform fully nonperturbative $O(a)$
improvement and aim at $O(2\%)$ computations of matrix elements
in quenched approximation.
To this end, we need to perform the calibrations
with accuracies much less than 1 \%.
We describe our strategy to tune the gauge and quark parameters
to this level in the quenched approximation.

\section{Anisotropic lattice quark action}
\label{sec:formulation}

Our heavy quark formulation basically follows the
Fermilab approach \cite{EKM97} but is formulated on the anisotropic
lattices \cite{Aniso01a,Ume01}.
The quark action is represented as
\begin{eqnarray}
 S_F &=& \sum_{x,y} \bar{\psi}(x) K(x,y) \psi(y),\\
 K(x,y) \!\!\!&=&\!\!\!
 \delta_{x,y}
   - \kappa_{\tau} \left[ \ \ (1-\gamma_4)U_4(x)\delta_{x+\hat{4},y} \right.
 \nonumber \\
 & &  \hspace{1cm}
      + \left. (1+\gamma_4)U_4^{\dag}(x-\hat{4})\delta_{x-\hat{4},y} \right]
 \nonumber \\
 & & \hspace{-0.2cm}
    -  \kappa_{\sigma} {\textstyle \sum_{i}}
         \left[ \ \ (r-\gamma_i) U_i(x) \delta_{x+\hat{i},y} \right.
 \nonumber \\
 & & \hspace{1cm}
     + \left. (r+\gamma_i)U_i^{\dag}(x-\hat{i})\delta_{x-\hat{i},y} \right]
 \nonumber \\
 & & \hspace{-0.2cm}
    -  \kappa_{\sigma} c_E
             {\textstyle \sum_{i}} \sigma_{4i}F_{4i}(x)\delta_{x,y}
 \nonumber \\
 & & \hspace{-0.2cm}
    - r \kappa_{\sigma} c_B
             {\textstyle \sum_{i>j}} \sigma_{ij}F_{ij}(x)\delta_{x,y},
 \label{eq:action}
\end{eqnarray}
where $\kappa_{\sigma}$ and  $\kappa_{\tau}$ 
are the spatial and temporal hopping parameters, $r$ the spatial
Wilson parameter and  $c_E$ and $c_B$  the clover coefficients.
For a given $\kappa_\sigma$, in principle, the 
four parameters $\gamma_F \equiv \kappa_{\tau}/\kappa_{\sigma}$,
$r$, $c_E$ and $c_B$ should be tuned so that the Lorentz invariance
holds up to discretization errors of $O(a^2)$.
We can set $r=1/\xi$ without loss of generality
\cite{Aniso01a,EKM97}.

In the first stage of this work,
we tuned only the bare anisotropy $\gamma_F$ nonperturbatively
and applied the tree-level tadpole-improvement %\cite{LM93}
to $c_E$ and $c_B$:
$c_E= 1/u_{\sigma} u_{\tau}^2$, $c_B = 1/u_{\sigma}^3$.
In the second stage, however, we need to perform the
nonperturbative tuning of all the three parameters $\gamma_F$,
$c_E$, and $c_B$, as well as the parameters of the operators
which appear in the matrix elements.

\section{Summary of the first stage results}
\label{sec:stage1}

In the first stage of this work, we have obtained the following
results in the quenched approximation.

{\it One-loop perturbative calculation} \cite{Aniso01a}:
Renormalization factors of heavy-light bilinears and
quark rest mass at
$m_Q a_\sigma \sim 1$, $m_Q\ll a_\tau^{-1}$ are calculated
in the one-loop perturbation theory.
The $m_Qa_\tau$ dependence of the coefficients are
well approximated with linear form
and this means that the quark mass dependence can be controlled.

{\it Numerical simulation in the quenched approximation}
\cite{Aniso01b}:
The mass dependent tuning of $\gamma_F$ is performed
with meson dispersion relation.
Quark mass dependence is small for $m_q a_\tau \ll 1$
and well fitted to a linear form in $(m_q a_\tau)^2$.

{\it Test of relativity relation} \cite{Aniso02a}:
Heavy-light meson dispersion relation is computed
with $\gamma_F=\gamma_F(m_q=0)$.
The relativity relation well holds for the region
$m_qa_\tau \ll 1$ while $m_q a_\sigma \simgt 1$.

{\it Application to the decay constant} \cite{Aniso02_fD}.
Around the charm quark mass, the heavy-light decay constant
is calculated.
The result is consistent with previous works. % \cite{Ryan02}.
This implies that our program successfully works at least
for computations at $O(10\%)$ accuracy.

These results are encouraging for further development
along this direction.

\section{Strategy for the second stage}
\label{sec:stage2}

The goal of the second stage is to perform
the quenched computations of heavy-light matrix elements
within a few percent uncertainties.
To achieve this precision, the calibration must
be performed to the level of accuracy much less than one percent
both for the gauge and quark fields.

\subsection{Calibration of gauge field}
\label{sec:calib_gauge}

In the quenched approximation, the calibration of gauge
field can be performed independently of the quark field.
The elaborated work by Klassen \cite{Kla98},
the $O(1\%)$ level calibration for the Wilson action,
is no longer enough for the present purposes.
For more precise calibration of gauge field,
we need to measure the static quark potential very
accurately.
For this purpose, we adopt the L\"uscher-Weisz noise reduction
technique \cite{LW01}.

We define the renormalized anisotropy $\xi_G$ through
the hadronic radius $r_0$ measured in the spatial and the temporal
directions.
Since we carry out the continuum extrapolation in terms of the lattice
scale set by $r_0$, the renormalized anisotropy is kept
fixed during the extrapolation.
This procedure prevents the systematic uncertainties
due to the anisotropy from remaining in the continuum limit.

\begin{figure}[tb]
\vspace*{-0.2cm}
\psfig{file=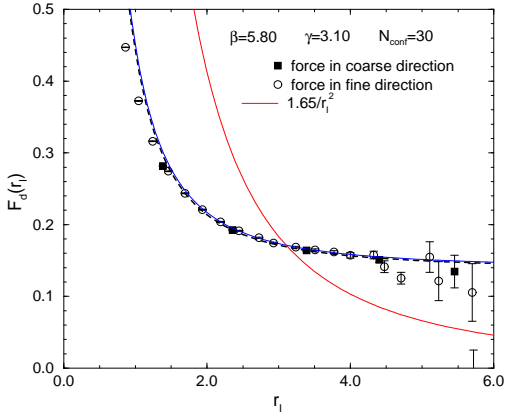,width=\figwidth}
\vspace{-1.4cm}
\caption{Determination of gauge field anisotropy.}
\label{fig:gauge1}
\vspace{-0.8cm}
\end{figure}

\begin{figure}[tb]
\vspace*{-0.2cm}
\psfig{file=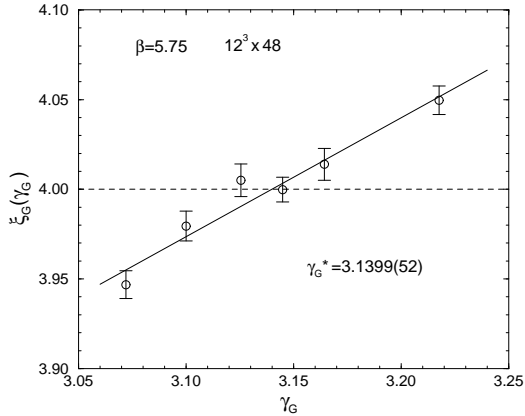,width=\figwidth}
\vspace{-1.4cm}
\caption{Calibration of gauge field at $\beta=5.75$.}
\label{fig:gauge2}
\vspace{-0.4cm}
\end{figure}

Figure~\ref{fig:gauge1} shows a result of static
quark potential at $\beta=5.8$ and $\gamma_G=3.10$:
the displayed data are the force between static quarks separated
in the spatial (coarse) and temporal (fine) directions.
The values of $r_0$, which is defined through the relation
$r_0^2 F(r_0)=1.65$, are determined within 0.2\%
statistical errors.
Fig.~\ref{fig:gauge2} exhibits the determination of $\gamma_G^*$,
which satisfies $\xi_G(\gamma_G^*)=\xi=4$, at $\beta=5.75$.
A linear $\chi^2$ fit gives $\gamma_G^*=3.1399(52)$,
which satisfies the required accuracy.
The calibration in wider range of $\beta$ at $\xi=4$
is in progress.

\subsection{Calibration of quark field}
\label{sec:quark_gauge}

We need to calibrate the parameters in the action,
$\gamma_F$, $c_E$, $c_B$, to the level which enables
computations of matrix elements within a few percent accuracy.
We also need to perform the nonperturbative renormalization
of the operators such as the heavy-light axial current.
The nonperturbative renormalization technique
\cite{NPR1,NPR2,NPR3} is one of the most powerful methods to 
perform such a program.
Since our expectation is that the result of tuning near
the massless limit can be applied to the heavy quark region
if $m_Qa_\tau\ll 1$ holds,
the technique can be applied with a little modification
in accord with the anisotropic lattice.

The Schr\"odinger functional method can be applied
to the anisotropic lattice in a straightforward manner
if the fine direction is realized as the temporal axis.
To examine the feasibility of the method
on anisotropic lattice, we perform the tree level analysis
in nonzero background field along Ref.~\cite{NPR2}.
Requiring the PCAC relation up to $O(a^2)$,
the tree level relations
$\xi/\gamma_F^{(0)}=1$ and $c_E^{(0)}=1$ are reproduced.
In this setting, $c_E$ can be tuned with sufficient
accuracy, while $c_B$ seems not.
The calibration of $\gamma_F$ may also be insufficient,
since it is the parameter to be tuned most precisely.

To tune all the required parameters to a sufficient level,
we perform the calibration of $\gamma_F$, $c_E$, $c_B$,
and the renormalization coefficients of the axial current along
the following steps.
(1) Tuning of $c_E$ by Schr\"odinger functional method
(and $\gamma_F$ if sufficient accuracy is accessible).
(2) Calibration of $\gamma_F$ and $c_B$ by requiring
 the physical isotropy conditions for $m_{PS}$ and $m_V$
 in the coarse and fine directions on lattices
 with $T$,$L$ $\simgt$ 2 fm.
(3) Determination of $\kappa_c$ and the renormalization coefficients
of the axial current by Schr\"odinger functional method.
(4) Finally, several checks are necessary.
We need to verify that the systematic errors are under control
by calculating the hadron spectra and the dispersion relations
and by taking the continuum limit.
It is also necessary to verify that the tuned parameters
in the massless limit is also available in the heavy quark mass
region.

The numerical simulation along this program is in progress.

%\bigskip

%H.~M. and T.~U. are supported by JSPS for Young Scientists.
%T.~O. is supported by the Grant-in-Aid of the Ministry
%of Education No. 12640279.
%The simulation has been done on
%NEC SX-5 at Research Center for Nuclear Physics, Osaka University and
%Hitachi SR8000 at KEK (High Energy Accelerator Research Organization).

\end{document}